\documentstyle[aps,prabib,epsfig]{revtex}
\draft

\def\vep{\varepsilon}
\def\vec#1{{\rm\bf #1}}

\def\matr#1{\underline{\underline{{\bbox{#1}}}}}
\def\lz{\ell_{\parallel}}
\def\lp{\ell_{\bot}}
\def\bea{\begin{eqnarray}}
\def\eea{\end{eqnarray}}

\begin{document}
\title{Electro-Mechanical Fredericks Effects in Nematic Gels}

\author{E. M. Terentjev, M. Warner}
\address{Cavendish Laboratory, University of Cambridge, Madingley
Road, Cambridge CB3 0HE, U.K.}
\author{R.B. Meyer and J. Yamamoto\footnote{Permanent address:
Institute of Industrial Science, University of Tokyo, Minato-ku, Tokyo 106, Japan}}
\address{The Martin
Fisher School of Physics, Brandeis University, Waltham, MA 02454-9110 USA}

\date{\today}
\maketitle
\begin{abstract}
The solid nematic equivalent of the Fredericks transition is found
to depend on a critical field rather than a critical voltage as in
the classical case.  This arises because director anchoring is
principally to the solid rubbery matrix of the nematic gel rather
than to the sample surfaces.  Moreover, above the threshold field,
we find a competition between quartic (soft) and conventional
harmonic elasticity which dictates the director response.  By
including a small degree of initial director misorientation, the
calculated field variation of optical anisotropy agrees well with
the conoscopy measurements of Chang {\it et al} (Phys.~Rev.~{\bf
E56}, 595 (1997)) of the electro-optical response of nematic gels.
\end{abstract}
\section{Introduction}\label{S1}

In nematic elastomers and gels, liquid crystal ordering and the
nematic director are coupled to the mechanical  degrees of
freedom.  Both symmetric shear strains and antisymmetric
deformations couple to the director.  Since the macroscopic shape
of the polymer network naturally depends on the chain anisotropy
direction, the coupling gives rise to some remarkable features of
the elasticity: shape can change spontaneously by well over 50\%
on changing temperature through the clearing point, shears  can
induce director rotation, and some shear deformations cost no
change in energy in the ideal case.  We call this last effect
``soft elasticity''.  It is  unique to nematic networks and other
elastic solids where a non-elastic internal degree of freedom is
coupled to the mechanical degrees of freedom.  The  additional
symmetries leading to softness were predicted phenomenologically
by Golubovic and Lubensky \cite{GL} and from  statistical
mechanics of large deformations by Bladon {\it et al}
\cite{BTWsoft}.  Depending  on the thermo-mechanical history,
nematic elastomers have been shown to demonstrate either extreme
softness or substantial deviations from soft response \cite{KF}.
Systems with such residual resistance  to deformation have been
called semi-soft \cite{review}.  They are qualitatively like soft
elastomers in that the same  particular modes of non-trivial
deformation cost a very small  elastic energy, compared to other
non-specific deformations.

Conventional nematics couple strongly to electric fields
($\vec{E}$), the director aligning parallel to the field in the
case of positive dielectric anisotropy $\Delta \vep$.  When the
director is anchored at boundaries and the sample is uniformly
aligned perpendicular to the direction to be taken by the electric
field, there is a transition (Fredericks) at a finite  critical
voltage, $V_F$, above which the director is deflected by the
field.  The electric Fredericks effect and its analogues are the
basis of most liquid crystalline displays.    Director rotation to
lower the dielectric energy in the bulk becomes non-uniform if the
boundary anchoring of the director is to be respected.  In this
case the Frank elastic energy density $\frac{1}{2} K (\nabla
\vec{n})^2$ opposes the field effect, where $K$ is a Frank
constant and $\vec{n}$ is the nematic director. For a cell of
thickness $d$ with strong anchoring on the boundaries the Frank
energy is of order $\frac{1}{2} K (\nabla \vec{n})^2 d \sim
\frac{1}{d}K$ per  unit area of cell (where $\nabla \vec{n} \sim
\frac{1}{d}$, crudely).   The corresponding electrical  energy per
unit area is, roughly, $\frac{1}{2}\vep_o \Delta \vep E^2 d$.
These are comparable for a field  of $E_F \sim
\frac{1}{d}\sqrt{\frac{K}{\vep_o \Delta \vep}}$ or an applied {\it
voltage}  $V_F \equiv E_F d =  \sqrt{K/ \vep_o \Delta \vep}$,
where the Fredericks transition occurs.

The response of the nematic director to applied electric fields in
nematic elastomers and gels raises questions about the role of the
elastic properties of the polymer network in  limiting the director
response.  Likewise, the reorientation of the director by an
external field can lead to elastic deformations.

Nematic elastomers have been seen to respond to modest electric
fields \cite{Efields} by changing their shape when they are
geometrically unconstrained.  At first sight this is unexpected
since the energy scale for shape change in a solid is of order
$\mu$, the shear modulus.  Balancing this with the electrical
energy, the characteristic fields  required to change shape should
be $E\sim \left(\frac{\mu}{ \vep_o \Delta \vep}\right)^{1/2} \sim
10^7{\rm V /  m}$ for a typical rubber modulus $\mu \sim 10^5 {\rm
J/m^3}$ and a substantial dielectric  anisotropy $\Delta \vep
\sim6$. This is a large field, whereas shape changes were observed
for  modest electric fields. However, because nematic elastomers
can suffer certain shape changes with  little or no energy cost,
which can potentially be achieved most easily in a completely
unrestricted  sample, such observations add weight to the concept
of soft mechanical response.

More recent experiments of Chang {\it et al} \cite{rbm} show an
electro-optical response in a nematic gel in the usual constrained
Fredericks geometry.   They observe the director response by the
method of optical conoscopy, and find an elevated threshold {\it
field} rather than voltage for director reorientation.  They
analyze their data in the classical Fredericks manner.  However,
to explain the threshold field, they invoke both a bulk anchoring
mechanism and an intrinsic, material-defined length scale for
gradients of the director, rather than the sample thickness, which
sets the length scale for a simple nematic. Given that finite fields
were required, Chang  {\it et al}
were not observing pure soft elasticity of the gel.  It is not
clear that the nematic gel can react softly in the Fredericks
geometry, since the delicate combination of shears required to
accommodate the director at no cost of elastic energy is
frustrated by the cell electrodes, to which the nematic solid must
conform.

At this point it is important to recall another physical system,
prepared in a way similar to the swollen nematic gels of Chang
{\it et al} \cite{rbm}. Polymer-stabilized liquid crystals (PSLC)
\cite{kent} are formed from a mixture of common nematic and
polymerizable monomers. On polymerization, complete phase
separation occurs and, when the polymerization occurs in an
aligned geometry, the resulting polymer fibrilles preserve the
direction of anisotropy by creating a large amount of internal
boundary.  Thus, the phase-separated polymer mesh provides both a
bulk anchoring mechanism and an intrinsic small (micrometer size)
length scale for director response. However, the response time for
director reorientation  for both field-on and off cases is fast
(which is one of the reasons for PSLC use). The compounds used by
Chang {\it et al} were highly miscible and had a slow response
dynamics, characteristic of nematic gels. We assume, therefore,
that their experiment dealt with an aligned homogeneous gel,
composed of cross-linked single molecular strands of nematic side
chain acrylate polymer, swollen by a solvent of similar mesogenic
molecules, rather than a  PSLC system.

In this paper we reconsider the experimental observations of Chang
{\it et al}.  We seek to explain the new electro-optical
transition in terms of a homogeneous response of the director,
rather than invoking a new small length scale.  The elevated field
threshold and the director response above threshold are accounted
for by the coupling of the director to the elastic polymer
network.  This analysis is partially supported by some new
experimental observations indicating that the samples respond to
the electric field on a scale comparable to the sample thickness,
rather than on a much smaller scale.  To summarize this new
analysis, we find that the free energy density for director
rotation through angle $\theta$ in a nematic elastomer is
schematically represented by
\begin{eqnarray}
f \approx \frac{1}{2} K \left( \frac{d \theta }{d z}\right)^2 -
\frac{1}{2} \vep_o \Delta \vep E^2 \sin^2\theta + {\mu} \left(
\frac{A}{2}\sin^2\theta + \frac{1}{4}\sin^4\theta\right) + \dots\; .
\end{eqnarray}
The Frank and the dielectric terms are as in classical nematics.
The rubber-elastic ${\mu}$ terms are the resistance the gel
network presents to director rotation, only the quartic term being
present in the ideal soft case.  The parameter  $A$ measures
semi-softness, that is the residual harmonic elastic resistance.
In elastomers and gels, the director has a bulk as well as  surface
anchoring -- there is an elastic resistance even when directors
are rotated  uniformly (equivalent to a massive, finite-energy
$q\rightarrow 0$ normal mode). One sees  that the uniform system
would rotate in response to an electric field only when the
threshold $E^* \sim \left(\frac{A\mu}{ \vep_o \Delta
\vep}\right)^{1/2}$ is exceeded, a characteristic field rather
than  voltage because of the bulk anchoring.  Additional effects
arise from gradients of $\vec{n}$  in order to satisfy the
boundary conditions and from the necessity for the sample to
preserve  its overall mechanical shape, but the main effect is
captured by the above argument.  We shall describe both effects in
this paper, first addressing the response of a uniform system  and
then examining the role of constraints and non-uniform
deformations.

\section{The electro-elastic-nematic energy}\label{S2}

\begin{figure}[h]
\centerline{\epsfig{file=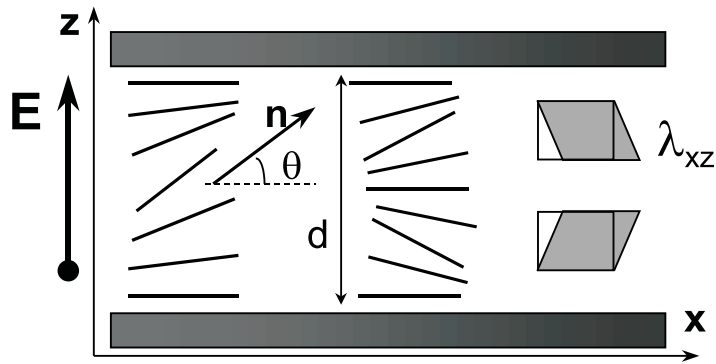,height=5cm}} \vspace{0.5cm}
\caption{ Field-induced director rotation in a conventional
nematic (left) and nematic elastomer (middle) anchored at the
surfaces $z=0$ and $z=d$, with an electric field $E$ applied
across the cell.  The shear strain $\lambda_{xz}$ accompanying the
director rotation in nematic elastomers is shown on the right. The
conventional Fredericks effect has one half wavelength of director
rotation between the plates, while the solid nematic Fredericks
effect has the full wavelength.} \label{fig1}
\end{figure}

We start with a model of a macroscopically uniform sample of
nematic elastomer in which the director is aligned parallel to the
plane of a dielectric cell. When subjected to an electric field
perpendicular to the cell plane (assuming positive anisotropy
$\Delta \vep$), the director is forced to rotate upwards towards
homeotropic alignment as in the ``traditional'' Fredericks effect.
We  assume for simplicity that the director rotates in the $x$-$z$
plane and is parameterized by an angle $\theta$, which is a
function of coordinate $z$ only, $\theta=\theta(z)$, but not of
$x$, just as in a classical Fredericks transition (see Figure
\ref{fig1}). Mechanical constraints of (i) rubber
incompressibility and (ii) elastic compatibility then impose
strong limitations on the number and types of possible elastic deformations.
Since the director orientation is coupled to the elastic network,
director rotation in the $x$-$z$ plane causes deformations
$\lambda_{xx}$ and $\lambda_{zz}$ (and $\lambda_{yy}$ from the
incompressibility constraint $Det \, \matr{\lambda}=1$), and
especially the shears $\lambda_{xz}$ and $\lambda_{zx}$. All these
components of strain are possibly functions of $z$. There are
restrictions due to mechanical (in fact, geometric) compatibility
\cite{landau}, which can be expressed in the following simple way.
Since the deformation (Cauchy strain) tensor is a derivative
$\lambda_{ij} =
\partial R_i/\partial R^o_j$ with $\vec{R}^o$ and $\vec{R}$ being
positions of a material point initially  and after deformation,
then one can obtain the second derivative in two possible ways:
\[
\frac{\partial \lambda_{ij}}{\partial R^o_k} \equiv
\frac{\partial^2R_i}{\partial R^o_j\partial R^o_k} = \frac{\partial
\lambda_{ik}}{\partial R^o_j}.\]

Making some elements $\lambda_{ij}$
functions of $z$ induces others to be functions of $x$.  For
instance, consider the component of shear $\lambda_{zx}(z)$: we
have $\partial \lambda_{zx}/\partial z=\partial
\lambda_{zz}/\partial x$, {\it i.e.} the extension $\lambda_{zz}$
must be a function of $x$, which we assumed prohibited. Hence
$\lambda_{zx}=0$. Similarly, assuming the local extension
$\lambda_{xx}(z)$ leads to $\partial \lambda_{xx}/\partial
z=\partial \lambda_{xz}/\partial x$, {\it i.e.} $\lambda_{zx}(x)$,
another prohibited $x$-dependence. Hence $\lambda_{xx}={\rm
const}$. However, incompressibility demands that $\lambda_{zz}
\lambda_{xx} \lambda_{yy}=1$.  Therefore, the constant extensional
strains in the cell plane lead to the conclusion that
$\lambda_{zz}={\rm const}$ as well. Analyzing the strain tensor,
we come to the conclusion that in a constrained cell geometry
depicted in Figure \ref{fig1} the only possible elastic
deformation is the shear $\lambda_{xz}$, a possible function of
$z$. Thence the overall strain tensor is
\bea
\matr{\lambda}=\left( \begin{array}{ccc} 1 & 0 & \lambda_{xz}
\\ 0 & 1 & 0 \\ 0 & 0 & 1 \end{array} \right). \nonumber
\eea
Nematic rubber elasticity depends on the anisotropic
step-length tensor $\ell_{ij}= \lp \delta_{ij} + (\lz-\lp) n_i
n_j$ of the network's polymer chains. Before deformation the
initial director is aligned along the cell, $n_x=1$, and the
corresponding step-length tensor is called $\matr{\ell}_o$. When
the director rotates, it can be parameterized by the angle
$\theta(z)$ via $n_x=\cos \theta ; \ n_z=\sin \theta$ and the
corresponding step-length tensor $\matr{\ell}$ is a function of
$\theta$. It is useful to introduce a ratio $r=\lz/\lp$, the
measure of chain anisotropy. Main-chain liquid crystal polymers
may have $r\geq 10$. The siloxane side-chain polymers used by H.
Finkelmann \cite{KF} have $r \sim 2-3$. The acrylates used in the
experiments of G.R. Mitchell \cite{mitch} and of Chang {\it et al}
\cite{rbm} have a much
lower anisotropy of the backbone, $r \leq 1.3 \ - \ 1.5$.

The cell boundaries impose additional nematic and mechanical
constraints: (i) as in the classical Fredericks effect, the
director should respect the anchoring conditions,
$\theta(0)=\theta(d)=0$ and (ii) no overall macroscopic
displacement along $x$ is possible, so the strain $\lambda_{xz}(z)
$ must have an integer number of oscillations (see the right of
Figure \ref{fig1}; this favors a full wavelength Fredericks
transition, in contrast with the half-wavelength Fredericks
transition of the ordinary nematic, left in the picture).

With all these preliminary restrictions and conditions, the full
free energy density of the system takes the form:
\bea
f=\frac{1}{2}\mu \ {\sf Tr}\left[ \matr{\ell}_o \cdot
\matr{\lambda}^T \cdot \matr{\ell}^{-1} \cdot \matr{\lambda}
\right] + \frac{1}{2}\mu \, A \, (\sin \theta - \lambda_{xz}\cos
\theta)^2  - \frac{1}{2} \vep_0 \Delta \vep
E^2 \sin^2 \theta + \frac{1}{2} K (\nabla \vec{n})^2 \label{F0}
\eea
The first term is the ideal nematic rubber elastic energy
density \cite{review} and can lead to the soft elastic response. The
second term is the appropriate non-ideal, ``semi-soft''
contribution observed in many elastomers and, in particular, in
those formed in a field-aligned monodomain state \cite{isabel}.
In elastomers $\mu$ gives the rubber energy scale,
essentially a shear modulus at small deformations. The
dimensionless parameter $A$ measures the (usually small)
semi-softness. Inserting the deformation $\matr{\lambda}
$ and the director rotation $\theta$ (through $\matr{\ell}$),
these two rubber-nematic terms take the form
\bea
f_{rub}=\frac{1}{2}\mu \, \left[ \frac{(r-1)^2}{r}\sin^2
\theta + \frac{r-1}{r} \, \sin 2\theta \, \lambda_{xz} + \frac{1 +
( r-1) \sin^2 \theta }{r}\lambda_{xz}^2 + A \, (\sin \theta -
\lambda_{xz}\cos \theta)^2   \right] \label{Frub}
\eea
where we henceforth absorb the constant $3\mu/2$, the energy of
the relaxed state, into $f$. The last
two terms in equation (\ref{F0}) are conventional for liquid
crystals. We adopt a simplified form of the dielectric term, with
the field $E$ rather than the displacement $D$. The reasons for
and the applicability of this simplification are discussed in the
Appendix.

\section{Analysis of uniform director rotation}  \label{S3}

Without any assumptions of small deformations or angles, the
optimal shear strain is given by the minimization of the free
energy density $f_{rub}(\theta,\lambda)$, Eq. (\ref{Frub}), with
respect to $\lambda_{xz}$ at a given $\theta$.  One obtains, after
some straightforward algebra: \bea \lambda_{xz}(z) &=& -
\frac{(r-A-1) \sin 2\theta  } {2\left[1+A r +(r-A-1)\sin^2 \theta
\right]} , \label{shear} \eea where the local director rotation
angle is, in fact, a function of $z$, $\theta=\theta(z)$.
Substituting this back into the free energy density gives an effective
free energy density, $f^*$, depending
only on the `liquid crystal' variable $\theta$.
\bea
f^*(\theta)  &=& \frac{1}{2}\mu \, \left\{\frac{K}{\mu} (\nabla
\vec{n})^2- \frac{ \vep_0 \Delta \vep E^2}{\mu} \sin^2
\theta  + \right.  \label{F1}  \\
&& + \left. \bigg[ \, \frac{(r-1)^2
\sin^4 \theta}{1+(r-1)\sin^2 \theta} + \frac{A \, r^2
\sin^2\theta}{(1+(r-1)\sin^2 \theta) (1+A \, r +(r-A-1)\sin^2
\theta)} \bigg]\right\}     \nonumber
\eea
The effect of the
underlying rubbery network is expressed by the term in square
brackets, $f^*_{rub}(\theta)$, an effective penalty for the
uniform director rotation imposed by the elastic gel network. This
expression is arranged in such a way that the contribution from
the ideal soft nematic rubber elasticity is separated from the
additional semi-soft part, the term proportional to the parameter
$A$.

One can notice that at small deformations, $\theta \ll 1$, only
the semi-soft correction contributes to the elastic energy, the
first term in brackets being proportional to the higher power
$\theta^4$ (appropriately expressing the softness of the ideal
rubber-nematic response). Therefore, the main threshold for the
uniform director rotation is determined by the semi-soft parameter
$A$, which controls the counter-torque to the electric field.
Note, that any effect of the field is scaled by the rubber modulus
factor, yielding a reduced dimensionless electric
field ${\cal E}=\left(\vep_0 \Delta \vep E^2/\mu
\right)^{1/2}$. In many cases this makes the response extremely
small. However, the highly swollen network used in the experiments
\cite{rbm} (up to 90\% solvent) may well have a much lower
modulus (as will be indeed confirmed in Section~\ref{S6}).
Also, as we shall see below, the scaling depends on
the important polymer anisotropy $(r-1)$ too, so that the materials
with small backbone chain anisotropy have, in fact, a stronger
response to electric fields.

In spite of the obvious non-linearity of the effective free
energy density (\ref{F1}), it is possible to solve  exactly for
the optimal director angle $\theta(E)$ by minimizing $ f^*(\theta)$.
The uniform, field-induced rotation of the director is
\bea
\sin^2\theta &=&  -\frac{A r+1}{r-Ar -1} + \frac{\sqrt{Ar+1}}
{\sqrt{r-1 - {\cal E}^2 }\sqrt{r-Ar-1}} \ ; \label{sin2}  \\
{\rm or} \ \ \ \theta^2 &\approx & \frac{{\cal E}_c^2 }{(r-1 -
{\cal E}_c^2)^2 } \left[{\cal E}/{\cal E}_c - 1\right]\; , \label{Te2}
\eea
where the
second expression shows the variation of angle just above the bulk
threshold.  The appropriate reduced variable $x={\cal E}/{\cal
E}_c - 1$ is used near the threshold field, ${\cal E}_c$.  The
value of this threshold is:
\bea
{\cal E}_c^2 = \frac{Ar^2}{Ar+1}\;,   \label{Ec}
\eea
and is  determined by both
the semi-softness parameter ({\it i.e.} the director locking in
the bulk) and the polymer chain anisotropy $r$. The solution Eq.(\ref{sin2})
reaches its upper-bound at the field value
\bea
{\cal E}_u^2 =\frac{(r-1)^2(r+1)}{r^2} - A \frac{r-2-Ar}{r}, \label{Eu}
\eea
when $\sin \theta=1$ and the director rotation is complete. All this
is in contrast with a conventional Fredericks effect, where there
is no barrier for the uniform bulk director rotation and, if no
boundary anchoring were involved, the director would switch to
$\theta=\pi/2$ immediately at $E\neq 0$.

The conoscopy technique, used in experiments \cite{rbm}, reveals
the change in optical path length for extraordinary rays
traversing the sample $\delta= \frac{d}{\lambda}(n_e-n_{eff})$,
with $d$ the cell thickness, and $\lambda$ the wavelength of
light. The effective refractive index depends on the current
director orientation $\theta (E)$ (we assume uniform rotation in
this Section) \bea n_{eff}^2 = \frac{n_e^2n_o^2} {n_o^2 +
(n_e^2-n_o^2) \sin^2 \theta }. \label{neff} \eea Near the
threshold, when $\theta \ll 1$, the change in optical path is
simply $\delta ({\cal E}) \approx \frac{d}{\lambda}
\frac{n_e^2-n_o^2}{n_o^2}n_e\theta^2({\cal E})$. A plot of $\delta
({\cal E})$ for $d/\lambda \sim 200$, $n_e=1.74, \ n_o=1.54$, very
small semisoftness $A=0.001$ and increasing values of the chain
anisotropy $r=1.1, \ 1.5$ and $2.5$ is given in Figure 2(a).
Chains with higher anisotropy $r$ show a stronger resistance to an
external torque, as one can see from equation (\ref{Ec}). At fixed
parameter $A$, the director response to the field is stronger when
the chain anisotropy $r$ is reduced, as can be seen from the
(expanded) equation (\ref{Te2}).  However the underlying
resistance to rotation is because of the semi-soft, quadratic
($A$) terms in the rubber-nematic free energy, which is why the
bulk threshold scales like ${\cal E}_c^2 \sim A$.  Only at larger
rotations do the quartic terms (of order 1 rather than $A$)
dominate, accounting for the scaling of the saturation field
${\cal E}_u^2$.  Figure 2(b) shows the optical path $\delta ({\cal
E})$ at fixed chain anisotropy $r=1.5$ and increasing degree of
network semi-softness, $A=0.001, \ 0.02, \ 0.05, \ 0.1$. At small
angles $\delta \propto \theta^2 $ and hence $\delta \sim ({\cal E}
- {\cal E}_c)$ [see Eq.(\ref{Te2})]. Therefore increasing $A$ (and
thus ${\cal E}_c$) increases the linear slope of the plots after
the threshold, as seen in Figure 2 (b).

\begin{figure}[h]
\centerline{\epsfig{file=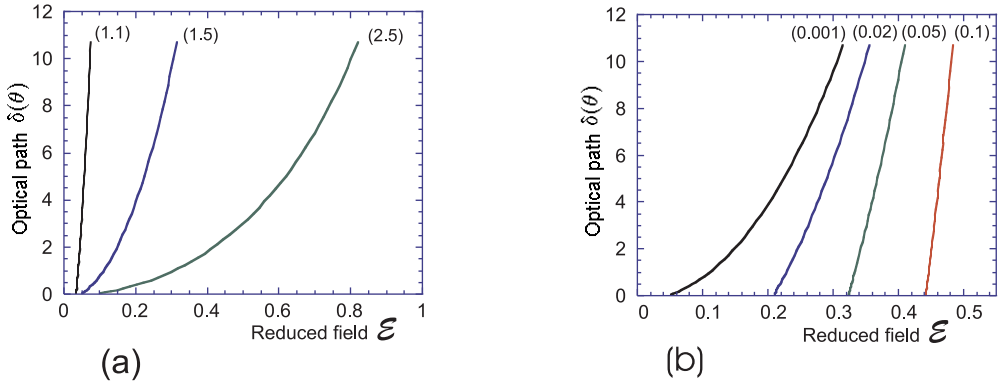,height=5cm}} \vspace{0.5cm}
\caption{ Change of extraordinary ray path length $\delta
(\theta)$ versus electric field.  (a) Small semi-softness $A=
0.001$ with increasing chain anisotropy $r = 1.1, 1.5, 2.5$ giving
less rapid growth of $\delta$ with ${\cal E}$. (b) At fixed
anisotropy $r=1.5$ with increasing $A = 0.001,\ 0.02,\ 0.05,\ 0.1$
showing an increase in the threshold field ${\cal E}_c$ with
increasing $A$. } \label{fig2}
\end{figure}

Figure 3 uses the further reduced field $x = {\cal E}/{\cal E}_c
-1$, and the optical pathlength scaled by $(r-1-{\cal
E}_c^2)^2/{\cal E}_c^2$, as suggested by equation (\ref{Te2}).
Deviations from the initial linear response are evident for high
threshold fields.  From equation (\ref{sin2}) for $\sin^2\theta$
we see there is a qualitative change in the field response for
${\cal E}_c^2 = r-1$, that is where the semi-softness $A =
(r-1)/r$. This condition is met either at low fields in networks
of low anisotropy, $r-1$ small, or for fairly strong
semi-softness, $A$ large.  At ${\cal E}_c^2 =r -1$ we have ${\cal
E}_u^2 = {\cal E}_c^2$.  There is no response to a field until it
reaches ${\cal E} = {\cal E}_c$, whereupon there is immediate,
discontinuous switching $\theta \rightarrow \pi/2$.

Clearly, the existence of a threshold in field, rather than
voltage, and quite a steep increase of the director rotation [and
consequently of the observed optical path $\delta ({\cal E})$] are
described well. However, there are two important aspects still
missing: (1) the analysis of boundary conditions and non-uniform
deformations, which leads to a Fredericks threshold
voltage in conventional nematics,
and (2) the experimentally observed small increase in
rotation before the principal threshold, a ``precursor foot''
which we shall associate with a small degree of random
misorientation in the initial director distribution.

\begin{figure}[h]
\centerline{\epsfig{file=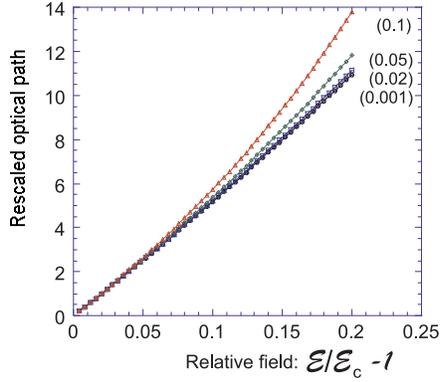,height=5cm}} \vspace{0.5cm}
\caption{ The conoscopic response for the conditions of figure
2(b).  The field is further reduced to $x = {\cal E}/{\cal E}_c
-1$, and optical pathlength is scaled by  $(r-1-{\cal
E}_c^2)^2/{\cal E}_c^2$, which collapses the plots of Fig. 2b for
various $A$ onto the single curve, Eq.(\ref{Te2}), in the region
of the transition, $x\ll 1$.} \label{fig3}
\end{figure}

\section{Critical analysis of the transition region}  \label{S4}

The analysis of the previous Section assumes that there is no
relevant spatial variation in either $\theta(z)$, or
$\lambda_{xz}(z)$: the solution for $\theta(E)$, equations
(\ref{sin2}) and (\ref{Te2}), has been obtained without taking into account the
director-gradient Frank elasticity. This is a reasonable
assumption at fields high above the threshold, where the texture
is coarsened and the bulk is capable of achieving its optimal
value of $\theta$, equation (\ref{sin2}), which we shall call
$\theta_0$ (similarly, for the conventional Fredericks effect
$\theta_0 \rightarrow \pi/2$ and the profile $\theta(z) $ coarsens
at high fields). Near the threshold one has to be more careful and
examine the effect of boundary conditions (which is the only
existing barrier for the conventional Fredericks effect).

Near the threshold we can safely assume $\theta \ll 1$ and the
free energy density takes the form, standard for elliptic-function
analysis (see Appendix)
\bea
\frac{1}{\mu} f^* \approx
\frac{1}{2}(\theta')^2
- \frac{1}{2}p \, \theta^2 + \frac{1}{4}q \, \theta^4,    \label{landau}
\eea
with the coordinate $z$ scaled
by the natural length scale $\xi=\sqrt{K/\mu}$. For a typical rubber modulus,
as in experiments \cite{isabel}, $\xi  \sim 10^{-8} \hbox{m}$.
Parameters $p$ and $q$ are obtained by expanding the
full free energy density (\ref{F1}):
\bea
p&=&{\cal E}^2 - \frac{A
\, r^2}{1 + Ar} , \label{params}  \\
q&\approx& 2\left[(r-1)^2 -
\frac{Ar^2}{1+Ar} \left(r-1 + \frac{r-1-A}{1+Ar}\right)\right].
\nonumber
\eea
In $q$ we have put ${\cal E}^2 \sim Ar^2/(1+Ar)$;
that is we are close to the threshold, and have neglected other
terms discussed in the Appendix.  The details of this type of
analysis are also given in the Appendix of paper \cite{isabel}:
the director cannot achieve its local optimal orientation
$\theta_0=\sqrt{p/q}$, being held by boundary conditions. Instead, the
director is only having a modulation of amplitude $\theta_m$
($<\theta_0$) across the cell. The resulting ``order parameter''
$\eta(E)=\theta_m/\theta_0$ increases from zero towards 1 as the
texture coarsens. The number of periods of the elliptic function
that describes the variation $\theta(z)$ is $\frac{d}{4
\sqrt{2}\xi {\sf K}} \sqrt{p(2-\eta^2})$ with ${\sf K} $ the
complete elliptic integral of argument $\eta^2/(2-\eta^2)$. The
minimal number of periods in our case is \underline{one} (in
contrast with the usual symmetric Fredericks effect, where it is a
\underline{half}), so one obtains the director rotation angle
$\theta_{m}$ as the solution $\eta(E)$ of
\bea
\frac{1}{\sqrt{2-\eta^2}}{\sf K}[\eta] &=&\frac{d \sqrt{p}}{4
\sqrt{2}\xi}\label{eta}
\\
&\equiv& \frac{1}{4\sqrt{2}}\frac{d}{\sqrt{K}}\sqrt{\vep_0 \Delta
\vep E^2-\mu \frac{A r^2}{1+Ar}} \qquad \ \ \left( \Rightarrow
\frac{1}{2\sqrt{2}}\sqrt{\frac{\vep_0 \Delta \vep}{K}} E \, d
\right)\; . \nonumber
\eea
The expression in parentheses is the
corresponding right-hand side for the conventional Fredericks
effect, with no underlying rubber elasticity $\mu$ and $A=0$ (and
with the half- rather than full-period of director oscillation
across the cell thickness); it depends on a voltage $V=E \, d$.
The real threshold is
shifted from that of the uniform analysis of last Section,
equation (\ref{Ec}), by the additional contribution from the
domain walls (which is the only contribution in the conventional
Fredericks effect): the solution of (\ref{eta}) first appears at
the minimal value of ${\sf K}=\pi/2$ at $\eta=0$, i.e. $p^*=(2\pi
\xi/d)^2$:
\bea
E^* \approx \sqrt{\left(\frac{\mu}{\vep_0 \Delta
\vep} \right) \frac{ A\, r^2}{1+Ar} + \left(
\frac{2\pi}{d}\right)^2 \frac{K}{\vep_0 \Delta \vep} } \qquad
\left[{\rm normally \ we \ expect \ \ } \frac{A \mu d^2}{K} \gg 1
\ \right]\; . \label{Estar}
\eea
One should note that the analysis
of the threshold itself is elementary and does not require the
investigation of the elliptic function.  In the usual way, we
could introduce a variation of wave vector $k$ $(= 2\pi/d)$
whereupon the expanded free energy density (\ref{landau}) is $f
\sim \frac{1}{2}(k^2-p)\theta^2 + \frac{1}{4} q\theta^4$.
Modulation starts when the $\theta^2$-coefficient becomes negative
for the first time, whence the threshold condition (\ref{Estar})
is recovered.

Returning to the elliptic analysis, near the threshold the
director rotation is given by $\eta^2= \frac{4}{3}(p-p^*)/p^*$,
describing the full-wavelength modulation with the amplitude
$\theta_{m} = \eta(E) \theta_0(E)$. The effect of this elliptic
analysis near the threshold rapidly disappears as the modulation
coarsens, the optimal value $\theta_0$ of the director rotation being
achieved in most of the sample and the $\delta(E)$ variation
becoming that of equation (\ref{neff}).

We said that one normally expects $(\mu d^2/K) \gg 1$, the
thickness-dependent term thus being irrelevant in the expression
for the threshold. As discussed below, effects observed in thin
samples suggest that the effective rubber modulus, $\mu A$, is low
in the studied material. Although this may in part be because
$\mu$ is small due to dilution by solvent, it clearly indicates that
$A$ is small too.  If the effect were simply due to the smallness
of $\mu$ alone, then the quartic terms would also be weakened and
the saturation field ${\cal E}_u$ would also be accessible.
Because $A$ is evidently small, it may be possible to see the
remaining $d$-dependence of the apparent threshold in experiment
and compare it with (\ref{Estar}).

\section{Oblique initial director}  \label{S5}

There seems to be no reason to assume a systematic obliqueness (pre-tilt)
in the initial director. However, the preparation of networks,
especially in thick cells, results in a residual polydomain
texture, as reported by Chang {\it et al.} and is a well-established
phenomenon in nematic gels. Even after orientation
by strong magnetic fields one should expect a small degree of
remaining random disorder, see \cite{fridrikh}. Assuming the
degree of this misalignment is small, it is straightforward to
make provisions for it and obtain the resulting quenched averages
of observable parameters, such as the director angle $\theta
({\cal E})$ and the optical path $\delta({\cal E})$.

\begin{figure}[h]
\centerline{\epsfig{file=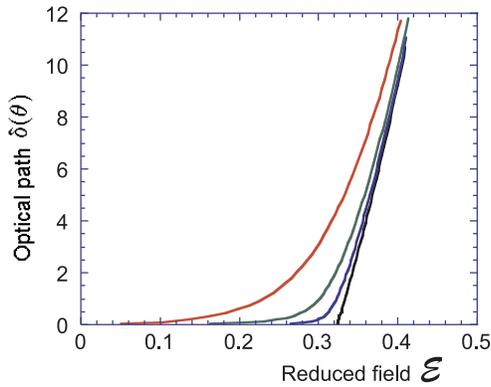,height=5cm}} \vspace{0.5cm}
\caption{ Conoscopic path length $\delta$ vs. reduced electric
field for backbone chain anisotropy $r=1.5$, semi-softness
parameter $A=0.05$ and increasing r.m.s. misalignment angle
$\alpha = 0, \ 0.01, \ 0.03, \ 0.05, \ 0.1$ radians.} \label{fig4}
\end{figure}

We do not go into details of this calculation, which involves some
heavy algebra. Instead, we plot the results for several values of
parameters.
The plots in Figure \ref{fig4} show the optical path $\delta({\cal
E})$ for selected values $r=1.5, \ A= 0.05$ and increasing r.m.s.
misalignment $\alpha=0, \ 0.01, \ 0.03,\ 0.05, \ 0.1$ (the last
value corresponds to $\alpha=5.7$ deg). The misalignment not only creates a
precursor foot before the main threshold, but also changes the
slope so that the extrapolated value of ${\cal E}$ appears
slightly reduced from $E^*$ in (\ref{Estar}). We emphasize that
theory and physical expectations suggest the obliqueness $\alpha$
should strongly depend on cell thickness and on the magnetic field
used during crosslinking.

\section{Experimental observations}  \label{S6}

We now apply the model of the previous sections to the
experimental observations of Chang {\it et al.} \cite{rbm}, in light of
further experiments performed recently.  The analysis of data
performed by Chang {\it et al.} includes a bulk anchoring effect
similar to that proposed here, which produces a threshold field,
rather than voltage.  However, they also appealed to an intrinsic
material length scale on the order of micrometers, which further
increases the observed threshold field and also sets the scale for
director gradients, which limit the director rotation above
threshold.  They reported observations of micrometer scale
speckles in the sample after polymerization in support of this
proposed length scale, but did not directly observe a periodic
pattern of director rotation at the micrometer scale in response
to the applied electric field.

We prepared a series of samples of different thicknesses according
the the methods of Chang {\it et al.}, and performed microscopic
observations during application of the field. In thick samples,
comparable to those of \cite{rbm}, we observed the same general elements of
response that they reported, but in addition, we were able to
observe small scale movements in the sample, particularly
displacements of microscopic particles and speckles in the sample.
We observed lateral displacements in the plane of the sample on
the order of a fraction of the sample thickness, and saw no
evidence of micrometer scale deformations in response to the
field.  Our observations are consistent with the deformation
sketched on the right in figure~\ref{fig1}, which involves a combination of
shear strains resulting in lateral displacement of material in the
midplane of the sample. The deformation proposed on the right in
Fig.~\ref{fig1} also satisfies the preservation of symmetry in the
conoscopic image reported by Chang {\it et al.}, in contrast to
the rotation of the optical axis that would accompany the classical
half-wavelength Fredericks transition.  Thus, the field response
proposed here is completely consistent with the general
observations reported by Chang {\it et al.} \cite{rbm}.  Moreover, as
described in the previous Section, the pretransitional response,
or rounding of the transition, observed by Chang {\it et al.} can
be similarly explained by a small random pre-tilt of the director.

\begin{figure}[h]
\centerline{\epsfig{file=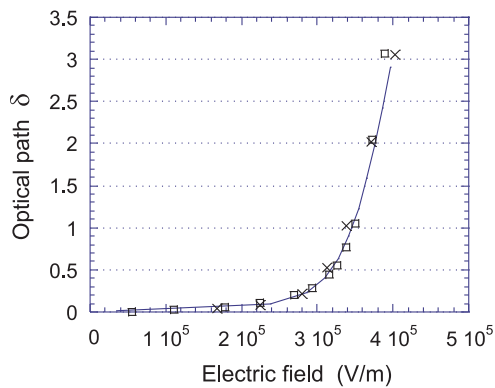,height=5cm}} \vspace{0.5cm}
\caption{Theoretical curve of $\delta(E)$, the conoscopic length,
against electric field for the present theory (solid line) for
thick samples (uniform director rotation) to the data of Chang
{\it et al.} \protect\cite{rbm}. For the 62 $\mu$m thick sample
(crosses) we have doubled the data values for the optical
pathlength $\delta$ to collapse the data from two samples onto the
same plot. For the 125 $\mu$m thick sample (squares) the data are
unscaled. This is the same scaling used in Fig. 1 of
\protect\cite{rbm}. } \label{fig5}
\end{figure}

Taking the published data of Chang {\it et al.}, and using the
model of Section~\ref{S3} appropriate for thick samples, along
with pretilt of the director, we present a fit of the present
theory to the data in figure~\ref{fig5}. We plot our theoretical
prediction of $\delta$ using $\theta$ from   \ref{sin2} and taking
the values $r$ = 1.95, $A$ = 0.06, $\mu = 1.6\times10^2$ J/m$^3$,
and $\alpha$ = 1.7 degrees for the network, in addition to the
sample characteristics specified by Chang {\it et al.}.  We see
that this theory can reproduce the experimental observations with
reasonable values of sample thickness etc. \cite{glitch}. The
values of semi-softness parameter $A$ and of the random local
pretilt angle $\alpha$ are sufficiently small not to contradict
the known facts about liquid crystalline gels. The value of
backbone anisotropy $r$, slightly higher than that reported for
solvent free polyacrylate elastomers ($\sim$ 1.44
\cite{mitch,backbone}), is likely due to the large amount of
nematic solvent present in the gel and increasing the nematic
field. The solvent (up to 90\%) can also explain the very small
value of rubber modulus $\mu$. In fact, it is mostly due to this
weakness of swollen elastic network that we can observe a
noticeable director rotation, the angle of which depends on a
ratio $E \left(\vep_o \Delta \vep / \mu \right)^{1/2}$. However,
in spite of this plausibility, independent measurements of these
parameters are needed to test the presented theory more
explicitly.

The value of the gel modulus $\mu$, used in the fitting above, is
substantially lower than that expected for ordinary rubbers. It
necessitates re-examination of the characteristic length scales in
our system. The natural nematic elastomer scale $\xi =
\sqrt{K/\mu}$ (see Eq.(\ref{landau}) and, more comprehensively,
the review \cite{review}) is now of the order $10^{-6}$m. This
scale gives the characteristic thickness of domain walls, two of
which are present in the full-wavelength Fredericks geometry, Fig.
1. Hence one expects that thin samples, with $d$ only of order of
several micrometers, should deviate from the description proposed
here: In thick samples, our electro-mechanical response is driven
by a reduction in electrical energy because of director rotation
toward the field with very little elastic penalty, since the
deformations are soft in the uniform regions above and below the
midplane. These regions are separated by a domain wall in the
middle, where the shear strain reverses sign.  When the sample
thickness is reduced, this wall and the two non-uniform regions
near the cell boundaries would start merging. This raises the
elastic energy of the full-wavelength (nematic elastomer-like)
Frederiks transition above that of the half-wavelength (liquid
nematic-like) one. Indeed, in thin samples (8 to 25 micrometers in
thickness), we observed the different response, almost identical
to the ordinary half-wavelength Fredericks transition.  This was
evident in the conoscopic observations, by the rotation of the
optical axis in response to the applied field, as in low-molar
mass nematic liquid crystals.

To satisfy the boundary conditions with this fully non-uniform director
rotation, the predominant shear strain in the bulk
of the sample must be zero.  The energy penalty and the counter-torque for
the action of electric field are then provided by Eq.(\ref{Frub}) with
$\lambda_{xz}=0$. This is the regime of ``couples without strains'',
with a bulk barrier for the director rotation, first proposed by de Gennes
\cite{degen}. More detailed
experimental measurements and theoretical analysis will be carried
out to explore this kind of transition in thin samples, and to
determine the thickness at which the crossover from thin to thick
behavior occurs.  This will provide a detailed understanding of
the combined director and shear strain structure within the boundary layers
and give an independent measure of the effective rubber modulus $\mu$ via
the characteristic length scale $\xi$.

Another aspect that needs to be re-examined for a system with very
low modulus $\mu$ is the threshold condition, Eq.(\ref{Estar}). In
order to predict the field threshold $E^*$, rather than the
voltage $V^*=E.d$, we need to maintain the inequality $A \mu d^2/K
\gg 1$. For the values used in fitting the experimental data in
Fig.5 (for the smaller $d\simeq 60 \, \mu\hbox{m}$) we still
obtain a ratio of order $3\times 10^{3}$ and the conclusions about
the threshold remain valid. However, this ratio will become of
order unity for $d\sim 10^{-6}$ m, where the ordinary voltage
threshold should again become relevant.

\section{Conclusion}  \label{S7}

In summary, we propose a model for the Fredericks transition in
nematic polymer gels.  Especially for thick samples, the polymer
network serves both to set the threshold field for the transition,
and to control the director response above threshold.  It effectively
replaces the role of the sample surfaces in the
ordinary nematic Fredericks transition.  We can reasonably explain
the experimental results of  Chang {\it et al.} without appealing
to a new internal length scale for director gradients, and this
element of our explanation is supported by qualitative
experimental observations of macroscopic lateral displacements in
response to the applied field.

Because two coupled physical fields, the director angle $\theta$
and the shear strain $\lambda_{xz}$, are at play, it is possible
that there are two different characteristic timescales for
relaxation. Depending upon the disparity of these, one could at
short timescales possibly see another decay regime before the
final, slowest exponential decay reported by Chang {\it et al.}.
This obtains naturally from the numerical solution of the
dynamical equations in $\theta $ and $\lambda_{xz}$ resulting from
Eq.(\ref{F0})  (see \cite{Texeira}).

We have begun to examine the remaining role of director and shear
strain gradients in the sample, and our experimental observations
suggest that this is important for understanding the transition in
thin samples. Much experimental work remains to be done to test various
aspects of our model.
Finally, it is interesting to compare these predictions for
nematic gels with the behavior of polymer-stabilized liquid
crystals. The key difference is the permanently fixed macroscopic
polymer skeleton of the PSLC system, in contrast with the mobile
molecular scale network of a gel. The detailed analysis of the
threshold, director evolution at higher fields and the role of
alignment conditions at preparation should distinguish between the
behavior of these two systems.  The dynamic behavior is especially
different, with gels responding very slowly, compared to PSLC's.

\begin{center}ACKNOWLEDGEMENTS \end{center}

This research was supported in part by the National Science
Foundation through grant DMR-9415656 at Brandeis university, and
by EPSRC in the UK.

\appendix
\section{ Analysis of dielectric problem }

\noindent In the classical electric Fredericks problem the
anisotropic dielectric fluid is contained between two parallel
plates kept at a relative voltage $V$ \cite{deJeu}. The analysis
can easily be generalized to the elastic-electric case.  One needs
to use the electric displacement vector $\vec{D} = (0,0,D)$, which
is a constant between the plates given no space charge ($div \,
\vec{D}=0$). The  field $E$ and the displacement $D$ are related
by $\vec{D} = \vep_o \underline{\underline{\vep}} \cdot \vec{E}$.
As all nematic uniaxial tensors, the matrix of dielectric
constants is expressed as $\vep_{ij} = \vep_{\bot}\delta_{ij} +
(\vep_{\parallel}-\vep_{\bot})n_in_j$, where the nematic director
\vec{n} is the principal axis. The voltage is, therefore, given by
the integration $\int E_z dz$: \bea V= \int_0^d dz
\frac{D}{\vep_o(\vep_\bot + \Delta\vep \sin^2\theta)},
\label{volt} \eea where $\vep_{zz} = \vep_{\bot} + \Delta \vep
\sin^2\theta$. The correct free energy, now per unit area of
plates, does not derive from equation (\ref{F0}), but from the
dielectric energy density $-(\vec{D}\cdot \vec{E})$: \bea F
&=&\int_0^d dz \left[\frac{1}{2} K
\left(\frac{d\theta}{dz}\right)^2 + \mu g(s) -
\frac{1}{2\vep_o}D^2 \frac{1}{\epsilon_{\bot} + \Delta\epsilon
\sin^2 \theta}\right] \label{a1}\\ &\equiv& \int_0^d dz
\left[\frac{1}{2} K \left(\frac{d\theta}{dz}\right)^2 + \mu g(s)
\right] \ - \ \frac{ \vep_o V^2}{2} \frac{1}{\int_0^d
\frac{dz}{\vep_{\bot} + \Delta\vep \sin^2 \theta}} \  , \nonumber
\eea where $\mu g(s)$ is the rubber-nematic part of free energy
density, given by equation (\ref{Frub}). The Euler-Lagrange
equation giving the $\theta (z)$ minimizing this free energy is:
\bea K\theta'' = \mu \frac{dg}{d\theta} +
\left(\frac{D^2}{2\vep_o}\right) \, \frac{d}{d\theta}
\left(\frac{1}{\vep_{\bot} + \Delta\vep \sin^2 \theta} \right)
\label{a2} \eea where the constant $D$ has to be determined from
equation (\ref{volt}) and is thus a functional of $\theta(z)$. The
boundary conditions are $\theta' = 0$ at $\theta = \theta_m$ if
there is a modulation with amplitude $\theta_m$. Then Eq.
(\ref{a2}) is easily integrated once to give: \bea
\left(\frac{d\theta}{d u}\right)^2 = 2\left(g(\sin^2
\theta)-g(\sin^2 \theta_m)\right) +
\frac{D^2}{\mu\vep_o}\left(\frac{1}{\vep_{\bot} + \Delta\vep
\sin^2 \theta}- \frac{1}{\vep_{\bot} + \Delta\vep \sin^2 \theta_m}
\right)\; , \label{a4} \eea where $u = z/\xi$ is the length
reduced by the rubber-nematic correlation length $\xi =
\sqrt{K/\mu}$.  The full problem is solved by further integrating
over 1/4 period and yields: \bea \int_0^{\theta_m}
\frac{d\theta}{\left[ \dots \right]} = \frac{d}{4\xi} \label{a5}
\eea where $\left[\dots\right]$ is the right hand side of Eq.
(\ref{a4}). The right hand side of Eq. (\ref{a5}) is a quarter
period of the modulation. Integrating Eq (\ref{a4}) to a general
$\theta$ at a general $z$, rather than to $\theta_m$ and $d/4$ as
in Eq. (\ref{a5}), gives $\theta (z)$ as an elliptic function,
which then finally determines $D$ via Eq. (\ref{volt}) and thereby
in Eq. (\ref{a5}) itself.

At the vicinity of the transition, all this analysis significantly
simplifies since $\sin^2\theta \leq \sin^2\theta_m \leq
\sin^2\theta_o  \ll 1$.  $D$ and $V$ are trivially related as $D =
\vep_o \vep_{\bot} V/d$.  The potential $g(s) +
\frac{D^2}{\vep_o}\frac{1}{\vep_{\bot} + \Delta\vep \sin^2
\theta}$ driving Eq. (\ref{a4}) simplifies to $-\frac{1}{2} p
\theta^2 + \frac{1}{4} q \theta^4$ with $p$ and $q$ given in the
equation (\ref{params}). The elliptic analysis then follows as if
from the simplified free energy Eq. (\ref{F0}).

In the uniform case considered in Section \ref{S3} the distinction
between $D$ and $E$ is easily handled too since Eq. (\ref{volt})
collapses to the obvious  $D = \vep_o ({\vep_{\bot} + \Delta\vep
\sin^2 \theta})V/d$ and the electrical part of equation (\ref{a1})
gives the $\theta$-dependent part of the electrical energy in
equation (\ref{F1}) (a constant part having been suppressed).


\end{document}